\documentclass{article}

\usepackage{arxiv}

\usepackage[utf8]{inputenc} 
\usepackage[T1]{fontenc}    
\usepackage{hyperref}       
\usepackage{url}            
\usepackage{booktabs}       
\usepackage{amsfonts}       
\usepackage{nicefrac}       
\usepackage{microtype}      
\usepackage{graphicx}
\usepackage{doi}
\usepackage{amssymb}
\usepackage{latexsym}
\usepackage{epsfig}
\usepackage{amsmath}
\usepackage{multirow}

\def\convinlaw{\stackrel{{\cal L}}{\Longrightarrow }}
\def\convinp{\stackrel{P}{\longrightarrow }}

\def\tends{\rightarrow}

\newtheorem{theorem}{Theorem}[section]

\newtheorem{proposition}{Proposition}[section]
\newtheorem{corollary}{Corollary}[section]
\newtheorem{Remark}{Remark}[section]

\newtheorem{algorithm}{Algorithm}[section]

\title{Local Quadratic Spectral and Covariance Matrix
Estimation  }

\author{ Tucker S. McElroy\\U.S. Census Bureau \\
  4600 Silver Hill Road, Washington, DC 20233\\
\texttt{tucker.s.mcelroy@census.gov} \And 
       Dimitris N. Politis \\Department of
Mathematics and Halicioglu Data Science Institute  \\
University of California at San Diego \\ La Jolla, CA 92093-0112, USA
      \texttt{dpolitis@ucsd.edu} }

\date{}

\hypersetup{
pdftitle={Local Quadratic Spectral and Covariance Matrix
Estimation},
pdfsubject={},
pdfauthor={Tucker S.~McElroy, Dimitris N.~Politis},
pdfkeywords={Flat-top lag-windows,  HAC Covariance  Estimation,  Kernel smoothing,
  Local Polynomials, Long-run variance, Sample mean},
}

\begin{document}
\maketitle

 \begin{abstract}  The problem of estimating  the spectral density matrix $f(w)$ of a multivariate
 time series is revisited with special focus on the 
frequencies $w=0$ and $w=\pi$. 
Recognizing that the entries of the  spectral density matrix 
 at these two boundary points are real-valued,  we propose a new estimator constructed 
 from a  local  polynomial regression of the real portion of the multivariate periodogram.
The case $w=0$ is of particular importance,  since $f(0)$ is associated 
with the large-sample
covariance matrix of the sample mean; hence,  estimating $f(0)$ is crucial in order to conduct any sort of statistical 
inference on the mean.  We explore the properties of the local polynomial
estimator through theory and simulations, and discuss an application
 to inflation and unemployment.
\end{abstract}



\section{Introduction}

Suppose $X_1,\ldots,X_n$ are observations from
the  (strictly)  stationary vector-valued 
sequence $\{X_t, t\in {\bf Z} \}$ having   mean $\mu = EX_t$, and 
well-defined autocovariance sequence 
\[
\gamma (h) =E  \left[ (X_{t+h} -\mu ) {(X_{t} -\mu ) }^{\prime} \right] ;
\]
  here  both $\mu$ and $\gamma (\cdot )$ are typically unknown.
   Let $m$ denote the dimension of $X_t$, and let $X_{t,k}$
denote the $k$th component of $X_t$ for $1 \leq k \leq m$.

 Under typical weak dependence conditions,  
 the   spectral density matrix evaluated at point $w\in  [-\pi, \pi]$ can be defined as 
 \[
 f(w)=   \frac{1}{2 \pi} \sum_{h=-\infty}^\infty e^{-iwh} \gamma (h) ; 
 \]
see Hannan (1970), Brillinger (1981), Rosenblatt (1985), 
Brockwell and Davis (1991),  and  Hamilton (1994).
 The spectral density matrix is $m \times m$---dimensional, and is Hermitian, i.e.,
 $ { f (w) }^{*} = f(w)$, where $*$ denotes the conjugate transpose operation.
  The Hermitian property ensures that $f(w)$ has $m$ real eigenvalues. 
 Moreover, the spectral density matrix $f(w)$ 
  is non-negative definite for each $w \in [-\pi, \pi]$, i.e., the eigenvalues of $f(w)$ are    non-negative.

Typical estimators of $\mu$ and $\gamma (h)$ are the sample mean
 $\bar X_n=n^{-1}\sum_{t=1}^{n} X_t$ and sample autocovariance
 $\hat \gamma (h)=n^{-1}\sum_{t=1}^{n-|h|} (X_{t+|h|}- \bar X_n) {(X_t- \bar X_n)}^{\prime} $ respectively;
  note that   $\hat \gamma (h)  $ is defined to be
 zero when  $|h|\geq n$.
 Under appropriate weak dependence and moment conditions---see e.g. Rosenblatt (1985)
or Wu (2005)--- the Central Limit Theorem (CLT) holds true,
namely 
\begin{equation}
\sqrt{n}(\bar X_n -\mu) \convinlaw N_m(0,\Omega) \ \ \mbox{as} \ \ n\to \infty
\label{eq.CLT}
\end{equation}
where $\convinlaw $ denotes convergence in law,  $N_m $ is the $m$--dimensional
normal distribution, and 
$\Omega =2\pi f(0)$.  It is apparent that accurate estimation of $f(0)$
is an important problem, as it is tantamount to accurate estimation
of $\Omega$, the large-sample covariance matrix
of the sample mean. It is also apparent that 
 $f(0)$ has all its elements real-valued (even the off-diagonal ones). 
 
Nonparametric estimation of $f(w)$ in the univariate case ($m=1$)  has its origins 
in the pioneering work of Bartlett  (1946),
Daniell (1946), and Parzen (1957, 1961). More recent developments
involve  using the so-called {\it flat-top} lag windows that ensure 
the fastest possible rate of convergence; see
Politis and Romano (1995),  Politis (2001, 2003), 
Paparoditis  and   Politis (2012), McElroy and   Politis (2014),
and McMurry and Politis (2015). 

In the multivariate case $m>1$, early developments can be 
found in Hannan (1970) and the references therein.
The subject was cast in the limelight in the econometrics literature via 
influential papers by 
Newey and West (1987, 1994), 
Andrews (1991), Andrews and Monahan (1992), and West (1997).
More recently, 
Politis (2011) constructed a non-negative definite estimator of
the spectral density matrix $f(w)$ based on the aforementioned 
flat-top  lag windows.
 
All the above works  developed  estimators of
 $f(w)$ that are valid for any $w \in [-\pi, \pi]$;
plugging in $w=0$, would yield
 the desired estimator of $f(0)$  (and of $\Omega$).
 Nevertheless, it was recently discovered that the case $w=0$ is special, and
deserves special treatment. Focusing on the 
univariate case,  McElroy and Politis (2022) argued that the even 
symmetry of  the (univariate) $f(w)$ 
makes   $w=0$ act as a boundary point in the nonparametric 
estimation of function $f(w)$; the same is true for the points $w=\pm \pi$. 
Hence, a local polynomial regression would be advantageous 
 on the boundary, and may yield an improvement
over the usual kernel smoothing; this is indeed true but one
has to use a local polynomial of order (at least) two,
since $f^\prime (0)=0$ due to the even symmetry of $f(w)$.
 
In the paper at hand, we investigate to what extent 
we can exploit the boundary effect at $w=0$ to construct an 
improved estimator of $f(0)$  (and of $\Omega$)
in the multivariate case. 
Section \ref{sec.method} lays the groundwork, while Section \ref{sec.methodLQ} 
presents the proposed methodology
that is based on local quadratic regression on the periodogram
ordinates. Section \ref{sec.asymptotic} presents   asymptotic consistency
 results, while
Section \ref{sec.practical} addresses some  practical aspects, including the important 
facet of ensuring a  non-negative definite matrix estimator.
 Section   \ref{sec.sims} contains  some  finite-sample simulations,
and  Section   \ref{sec.data} gives a real data application on 
the U.S. Consumer Price Index (CPI)  and Unemployment Rate (UR).

\section{Improved estimation of the spectral density matrix} 
\label{sec.method}

Let $f_{jk} (w) $ denote element $j,k$  of the
spectral density matrix $f(w)$. Note that 
   the diagonal entries of $f(w)$ correspond to univariate spectral densities.  In fact,  
 $f_{kk} (w) $ is the spectral density of the $k$th component time series of $\{ X_t \}$, i.e., the sequence $\{X_{t,k}, t\in {\bf Z} \}$. As such, $f_{kk} (w) $
is a real-valued, non-negative, and symmetric function for   $w \in [-\pi, \pi]$. 
However, the  off-diagonal elements of $f(w)$
are, in general, complex-valued.

Univariate spectral density estimation techniques can be applied to estimate the diagonal elements of $f(w)$.
 For example, one can use traditional kernel smoothing 
or local quadratic regression---see Ch.~9 of 
  McElroy and Politis (2022) for an overview and recent developments.  However, the off-diagonal
  entries correspond to the cross-spectral densities of two component time series, and the univariate
  techniques cannot be directly applied without some modification.  For one thing, the cross-spectral density
  can be complex-valued, and univariate spectral density estimation techniques rely on the target
  being real, non-negative, and an even function of the frequency $w$.

To elaborate, let $\gamma_{jk} (h)$ denote the $j,k$---th entry of $\gamma (h)$, 
i.e.,   
$\gamma_{jk} (h)=$ $\mbox{Cov} [ X_{t+h,j}, X_{t,k} ]$. 
 Then, the cross-spectral density between  components $j$ and $k$ is
 \[
   f_{jk} (w) = \frac{1}{ 2 \pi}  \sum_{h=-\infty}^\infty e^{-iwh} \gamma_{jk} (h) 
    =   \frac{1}{ 2 \pi}  \sum_{h=-\infty}^\infty \cos (w h) \gamma_{jk} (h) - i \,
     \frac{1}{ 2 \pi}   \sum_{h=-\infty}^\infty \sin (wh) \gamma_{jk} (h).
\]
 The first term on the right hand side is the real part  of the cross-spectrum  (sometimes called the cospectrum),
  whereas the second term is
  $i$ times the imaginary part (sometimes called the quadrature spectrum when multiplied by $-1$).
    In general this imaginary part is non-trivial, although it can be zero 
  (for example, if every $\gamma (h)$ is symmetric), and is always zero for $w =0, \pm \pi$ as already mentioned.
  The real part need not be non-negative---see Example 11.6.2 of Brockwell and Davis (1991)---but it is always   an even function of $w$.  
  
  The local quadratic regression technique of McElroy and Politis (2022) was proposed to give improved
  accuracy of univariate spectral densities near frequencies $w = 0, \pm \pi$.  Since the univariate spectral
  density has reflectional symmetry about the vertical axes at $w = 0, \pm \pi$,  one can develop a Taylor
  series expansion  in the neighborhood of such frequencies and obtain higher order accuracy;
this Taylor   series expansion has to be of order two (or higher) 
because the linear term vanishes due to the even symmetry.
 Our current proposal
  is to apply the same technique to the real part of the cross-spectral density for $w = 0, \pm \pi$.
  
Our concrete proposal in order to estimate the 
spectral density matrix $f(w)$ is given below: 

\begin{itemize}

\item [i.] To estimate   $f(w)$ for 
 $w \neq 0, \pm \pi$ we utilize the flat-top estimation methodology
detailed in Politis (2011); this includes a data-dependent bandwidth choice  as
well as a modification ensuring an estimator that is a non-negative definite matrix.

\item [ii.] If  $w = 0$ or $ \pm \pi$,  then the diagonal entries of $f(w)$ are estimated
  using the local quadratic regression technique of McElroy and Politis (2022);  the off-diagonal entries (i.e.,
  the $m^2- m$ cross-spectral densities) are estimated by applying the same local quadratic regression technique
  to the real part of the spectrum---since the imaginary part of the cross-spectal density is zero for $w = 0, \pm \pi$. 
A data-dependent bandwidth choice  and  a non-negative definite modification is also constructed. 
  \end{itemize}
 
Since part (i) is already well-known, we focus on part (ii)  in what follows.

\section{Local quadratic estimation 
 at the boundaries} 
\label{sec.methodLQ}
 
In the Introduction, it was alluded that weak dependence conditions
 are typically needed.
For our work, we will consider the following assumption:

\vskip .1in
\noindent
{\bf Assumption A$(p)$}: The spectral density matrix
 $f(w)$ is well-defined, and its $j,k$ element
$f_{jk} (w) $ is  $p$-times continuously differentiable for all real $w$, and 
for all $j,k$.
\vskip .1in
\noindent 
  From here on and throughout this paper, we will   assume {\it Assumption A($p$) holds for some integer  $p\geq 2$.}
 Note that    $f(w) $
 (and its derivatives) are periodic functions  with period $2\pi$; therefore, our 
attention focuses on $w\in [-\pi, \pi]$.
It is easy to see that if $\sum_{h=-\infty}^\infty |h|^p|\gamma_{jk} (h)|<\infty  $ for a non-negative  integer $p$,
  then $f_{jk} (w) $ is indeed  $p$-times continuously differentiable; see   Proposition 6.1.5 of McElroy and Politis (2020). 

The periodogram matrix is defined as $I(w)= ( 2\pi)^{-1} \sum_{h=-\infty}^\infty e^{-iwh} \widehat{ \gamma} (h) $; let $I_{jk} (w) $  denote its  $j,k$ element,
i.e.,  $I_{jk}(w)= ( 2\pi)^{-1} \sum_{h=-\infty}^\infty e^{-iwh} \widehat{ \gamma}_{jk} (h) $.
 A traditional spectral density estimator 
$\hat  f   (w)$ is obtained by means of smoothing the periodogram over the Fourier frequencies $w_s=2\pi s/n$, i.e., 
\begin{equation} \hat f  (w) = 
\frac{1}{2\pi n} \sum_{s\in J_n}    \Lambda_M(w_s) I(w+w_s)
\label{eq.lwseKapprox}
\end{equation}
where  the index set $J_n$ consists of $n$
consecutive integers as follows: 

$J_n=\{-\frac{n-1}{2},\ldots, 0,\ldots,  \frac{n-1}{2} \}$ if $n$ is odd, and

$J_n=\{-\frac{n}{2}+1,\ldots,0,\ldots,  \frac{n}{2} \}$  if $n$ is even.

In equation (\ref{eq.lwseKapprox}), 
 $\Lambda_M(w)= ( 2\pi)^{-1}  \sum_{s=-\infty}^\infty e^{iws}\lambda  (\frac{s}{M})$ 
is the so-called {\it spectral window} or {\it kernel}; it is based on
 $\lambda (x)$, which is a fixed even function---called the {\it lag window}.
Note that  
$\Lambda_M(w)$ becomes more concentrated around the origin 
as the parameter $M$ increases with the sample size $n$.
However, the kernel {\it bandwidth fraction}   defined as $\delta =M/n$ will
 typically be assumed to decrease as   $n$ increases.

Consider  element  $j,k$ of  eq. (\ref{eq.lwseKapprox}), i.e., 
\begin{equation} \hat f _{jk} (w) = 
\frac{1}{2\pi n} \sum_{s\in J_n}    \Lambda_M(w_s) I_{jk}(w+w_s).
\label{eq.lwseKapprox-jk}
\end{equation}
Focusing on the case where    $w=\theta$
with  $\theta= 0$ or $ \pm \pi$, recall that  $\hat f _{jk} (\theta)$
is real-valued; hence, we can use just the real part of $I_{jk}(w+w_s)$
  in  equation (\ref{eq.lwseKapprox-jk}), i.e., 
\begin{equation} \hat f _{jk} (\theta) = 
\frac{1}{2\pi n} \sum_{s\in J_n}    \Lambda_M(w_s) {\cal R } [I_{jk}(\theta +w_s)],
\label{eq.lwseKapprox-jkREAL}
\end{equation}
where ${\cal R } [\cdot]$ denotes the real part. 
Note that in case $j=k$, taking the real part is superfluous since 
$I_{kk}(w)$ is real; however, we will use eq. (\ref{eq.lwseKapprox-jkREAL}) 
 as it holds in general.

Under standard assumptions, 
  the periodogram ordinates $I_{jk}(w_s)$ for $s \in  J_n$ are approximately independent, 
  and   ${\cal R } [I_{jk}(w)]$ has even symmetry around
$\theta$. Hence,  equation (\ref{eq.lwseKapprox-jkREAL}) is tantamount
to applying (weighted) 
  local averaging on 
a nonparametric regression of ${\cal R } [I_{jk}(w)]$ as a function of $w$ 
with data points being the pairs $\left({\cal R } [I_{jk}(w_s)], w_s\right)$ for $s \in  J_n$.
 
Due to the even symmetry of $\Lambda_M$ and the real part of the cross periodogram,
 $w=0$ effectively acts as a {\it  boundary} point in the nonparametric regression of the 
periodogram ordinates ${\cal R } [I_{jk}(w)]$, and the same is true for the points $w=\pm \pi$ due to the periodicity
of $I(w) $. To see why, note that in such a case
equation (\ref{eq.lwseKapprox-jkREAL})  reduces to a one-sided sum
since the summands  less than $\theta$ are exactly the same 
as  the summands greater than $\theta$.
It is well known that local polynomial regression is 
better than plain local (weighted) averaging  
at boundary points. 
Therefore, we  propose local polynomial fitting of periodogram 
ordinates  instead of local averaging in order to estimate $f_{jk}(\theta) $.

Because of the even symmetry (and periodicity) of ${\cal R } [f_{jk} (w) ]$,
and the equally spaced data points  $w_s$,  local  linear fitting would  
be equivalent to local averaging when $w=\theta$. A local quadratic (without a linear term, due to the symmetry) 
can be fitted instead; 
for a  small value of the bandwidth   fraction $\delta \in (0,0.5)$,
 consider the Taylor approximation
\begin{equation}
 {\cal R } [f_{jk}(w) ] \approx a_0+b_0w^2  \  \  \mbox{for} \  \ w\in [-  2  \pi \delta,  2  \pi \delta],
\label{eq.loc-quad 0}
\end{equation}
 which  can be used to estimate $ f_{jk}(0)$, 
  as well as the  approximation
\begin{equation}
 {\cal R } [f_{jk}(w) ] \approx a_1+b_1(w-\pi) ^2   \  \  \mbox{for} \  \ w\in [\pi- 2  \pi \delta, \pi+ 2  \pi \delta]
\label{eq.loc-quad pi}
\end{equation}
to estimate $ f_{jk}(\pi) $.  Since $f_{jk} (-\pi) $=$f_{jk}( \pi) $, we do not have to address
the case $w=-\pi$ separately. 
\vskip .1in
Our proposal   then is summarized in the  following algorithm:
\begin{algorithm}
\label{algo.1}
\vskip .1in
\end{algorithm}

\begin{enumerate}
\item [(i)]
Let $\hat a_0$ and $\hat b_0$ 
be  the estimators of  $a_0$ and $b_0$  in eq. (\ref{eq.loc-quad 0}) based on  a
(possibly weighted)   regression of ${\cal R } [I_{jk}(w )]$ on 
a constant and a quadratic term $w^2$.
 The data to be used in this regression
are ${\cal R } [I_{jk}(w_1)], \ldots, {\cal R } [I_{jk}(w_M)] $ where $w_M$
 is the largest Fourier frequency less or equal to $2  \pi \delta$; 
since $w_M=2\pi M/n \leq 2  \pi \delta$, it follows that $M=[\delta n ]$ where $[\cdot]$ denotes the integer part.

\item [(ii)]
Similarly,  $\hat a_1$ and $\hat b_1$ 
are  the estimators of  $a_1$ and $b_1$  in eq. (\ref{eq.loc-quad pi}) based on a 
(possibly weighted)   regression of ${\cal R } [I_{jk}(w)]$ on 
a constant and a quadratic term  ${(w-\pi) }^2$.
  The data to be used in this regression
are the ${\cal R } [I_{jk}(w_s) ] $  with indices corresponding to the $M$ largest elements of the set $J_n$.

\item [(iii)] Finally,  $\tilde   f_{jk}(0)=\hat a_0$  is  the proposed new  estimator of $  f_{jk}(0)$,
and  $\tilde f_{jk}(\pi) =\hat a_1$ is  the proposed  estimator of $  f_{jk}(\pi)$.
 Furthermore, our preliminary estimator  of $f(\theta) $
  is the matrix $\tilde   f(\theta) $ with $j,k$ element $\tilde   f_{jk}(\theta) $;
this matrix estimator will be corrected towards positive definiteness in 
Section~\ref{sec.practical} in order to construct our improved estimator of
the long-run covariance matrix $\Omega$.
 
\end{enumerate}
 
\vskip .1in
  \noindent
If Assumption A($p$) holds with $p>2$, then 
a higher order polynomial (with only even powers) can also be used instead of the simple quadratics
(\ref{eq.loc-quad 0}) and (\ref{eq.loc-quad pi}), the goal again being to estimate the
respective intercept terms;  the details are straightforward and thus omitted. 
As will be shown in the next section, to achieve consistent estimation
in either of the above cases, 
 we would need $\delta \to 0$ but $\delta n\to \infty $ as $n\to \infty $, which is equivalent to 
$M\to \infty $ but with $M/n \to 0$.  

\section{Asymptotic results}
\label{sec.asymptotic}

For the purposes of deriving some asymptotic results, we formulate the additional
Assumption B below.

\vskip .1in
\noindent
{\bf Assumption B}: Suppose $\{ X_t \}$ is a strictly stationary time series that is either a linear process  (with 
 MA($\infty$)  coefficients that have square summable matrix norm, and inputs having a finite fourth moment
 for each component),  or 
   has autocumulant functions satisfying the summability condition (B1) of Taniguchi and Kakizawa (2000, p.55). 
\vskip .1in
\noindent 
Throughout this section, we will consider estimation of 
$    f_{jk}(\theta) $ for two fixed values of $j,k$.

\subsection{Fitting via Ordinary Least Squares (OLS)}
   
  The  Ordinary Least Squares (OLS)
 regression estimator  of $a_0$ and $b_0$  (or $a_1$ and $b_1$) 
 at one  of the boundary frequencies $\theta$ takes the form
 $ {( {\bf X}^{\prime} {\bf X} )}^{-1} {\bf X}^{\prime} {\bf Y}$,  where
 ${\bf X}$ has a first column of $M$ ones and a second column of $M$ values
 given by  ${(  \theta  - w_s)}^2$; 
   here,  the $w_s$ are the Fourier frequencies
   falling in the interval $[  \theta  - 2 \pi \delta,  \theta  + 2 \pi \delta] \cap ( 0, \pi]$.
   Also ${\bf Y}$ is  a vector with entries
${\cal R } [I_{jk}(w_s)]$,
 the periodogram evaluated at this
   set of Fourier frequencies.    To determine the Fourier frequencies that are used,
   note that by even symmetry at the boundaries we only need to focus on the interval
   $[0, \pi]$, and the scenario for $\theta = - \pi$ is the same as for $\theta = \pi$.
    Hence, for $\theta = 0$ we consider the set of $w_s \in (0, 2 \pi \delta]$,
    or $1 \leq s \leq M$ with $M =  [ \delta n]$.
    Note that  we do not use $s =0$ in the regression,  because  the periodogram
  we employ  is   mean-centered, implying that $I( 0) = 0$.
  For $\theta =   \pi$,  we consider the set of $w_s \in [  \theta  - 2 \pi \delta, \pi]$,
  or $ [ n /2] - M +1  \leq s \leq [n/2]$.   We can take the lower bound to 
  be  $[n/2 - n \delta]$  instead of  $[n/2] - M+1$  for purposes of asymptotic analysis.  
 Let $\convinp$ denote convergence in probability. 

 \begin{theorem} 
 \label{thm:OLS-consistent}
Assume Assumptions A(4) and B.  
Let  $ \tilde   f_{jk}(\theta)=  {[1, 0]}^{\prime} {( {\bf X}^{\prime} {\bf X} )}^{-1} {\bf X}^{\prime} {\bf Y}$.
Then,  $$  \tilde   f_{jk}(\theta) \convinp   f_{jk}(\theta) $$
as $M\to \infty$ and $n\to \infty$ but with
$M/n \to 0$.
\end{theorem} 
\vskip .1in
\noindent
{\bf  Proof:}    We focus on the case of $\theta = 0$, as $\theta = \pm \pi$ is similar with some
 notation changes.   Denote the entries of the symmetric matrix 
$M^{-1} {\bf X}^{\prime} {\bf X}$ by $c_0$ (upper left), $c_2$ (upper right), and
  $c_4$ (lower right).  Then $c_{2 \ell} = M^{-1} \sum_{s=1}^M w_s^{2 \ell}$.
  Letting $g(w) = (c_4 - c_2 w^2)/(c_4 - c_2^2)$, the OLS estimator 
  can be written 
$ \tilde{f}_{jk} (\theta) = M^{-1} \sum_{s=1}^M g(w_s) \, {\cal R } [I_{jk}(w_s)].$
 As a first step, note that we can replace the sample mean centering in $I_{jk}$
  (recall that it is the Fourier transform of the sample autocovariances, which are centered
  by the sample mean $\overline{X}_n$) by the true mean at the cost of $O_P (n^{-1})$ terms,
  following the arguments in Proposition 9.6.4 of McElroy and Politis (2020) -- adapted
  in a straightforward manner from the univariate peridogram to the cross-periodogram,
  using Assumption B.  Next, 
  define the matrix-valued function $\varphi (w) = .5 g(w) ( e_k e_j^{\prime} + e_j e_k^{\prime})$,
   where $e_j$ and $e_k$ are the $j$th and $k$th unit vectors of dimension $m$.  
   Because ${\cal R } [I_{jk}(w_s)] = .5 ( I_{jk} (w_s) + I_{kj} (w_s) )$, 
   it follows that $ \tilde{f}_{jk} (\theta)  = M^{-1} \sum_{s=1}^M \mbox{tr} ( \varphi (w_s) \, I (w_s) ) $.
   Recall that  $\delta = M/n$, so we can write
   $ \tilde{f}_{jk} (\theta)  = \delta^{-1}  n^{-1} \sum_{s=1}^n 1_{\{ [0, 2 \pi \delta ] \} } (w_s)
     \mbox{tr} ( \varphi (w_s) \, I (w_s) ) $;
   since ${\varphi (w)}^{\prime} = \varphi (w)$ and $I$ is centered by the true mean, 
      we can adapt the argument used in Lemma 3.1.1(i) of Taniguchi and Kakizawa (2000) for each fixed $\delta > 0$, 
      obtaining 
\[
 \lim_{\delta \tends 0} \limsup_{n \tends \infty} P  [  | \tilde{f}_{jk} (\theta) - \bar{f}^{\delta}_{jk} (\theta) | >
  \epsilon ] = 0
  \]
  for every $\epsilon > 0$,      where $\bar{f}^{\delta}_{jk} (\theta) = 
      \delta^{-1}  n^{-1} \sum_{s=1}^n 1_{\{ [0, 2 \pi \delta ] \} } (w_s)
     \mbox{tr} ( \varphi (w_s) \, f (w_s) ) $.   Since $c_{2 \ell} \tends {(2 \pi \delta)}^{2 \ell}/(2 \ell + 1)$
     as $n \tends \infty$       and $\theta = 0$,      it follows       that $\lim_{n \tends \infty} g( 2 \pi \delta x) =      (9-15 x^2)/4$.
     Then  $\lim_{n \tends \infty} \bar{f}^{\delta}_{jk} (\theta)
      = f^{\delta}_{jk} (\theta)$ for every $\delta > 0$, where
\[
 f^{\delta}_{jk} (\theta) = \int_0^1  (9-15 x^2) \, 
  ( f_{jk} ( 2 \pi \delta x ) + f_{kj} (2 \pi \delta x) )/8 \, dx.
  \]
Finally, using Assumption A(4), we have $f^{\delta}_{jk } (\theta) \tends f_{jk} (0) \, \int_0^1 (9 - 15 x^2)/4 \, dx = f_{jk} (\theta)$.
 Using 
 \[
  | \tilde{f}_{jk} (\theta)  - f_{jk} (\theta) | \leq
    | \tilde{f}_{jk} (\theta) - \bar{f}^{\delta}_{jk} (\theta) | + 
     | \bar{f}^{\delta}_{jk} (\theta) - {f}^{\delta}_{jk} (\theta) | + | {f}^{\delta}_{jk} (\theta) - f_{jk} (\theta) |
\]
 and the above estimates, we obtain the desired convergence in probability
  (following the argument of the proof of Theorem 25.5 of Billingsley (1995))
  and letting $\delta \tends 0$ as $n \tends \infty$.    q.e.d
  
  \vskip .13in

   Recall that the bandwidth fraction is $\delta =  M/n$,
 which is assumed to tend to zero in the above.
However,  a fixed bandwidth fraction calculation can be most informative, as  
    the pioneering work of Kiefer   and  Vogelsang   (2002, 2005) has shown; see also
   McElroy and Politis (2014). 
We now provide an asymptotic analysis of bias and variance in terms of a fixed $\delta $,
 which will help  develop an expression for the optimal bandwidth fraction. To   simplify the calculation, we will work under the
 assumption of a Gaussian time series.  
However, the Gaussian assumption is not unduly limiting here, as
the extra factors  associated with   potentially non-Gaussian 
data are asymptotically negligible; see McElroy and Politis (2022)
for a full discussion.    Define
\begin{eqnarray}
\label{eq:Fp}
  & F_{2p} & = M^{-1} \sum_{s=1}^M {(\theta - w_s)}^{2p}  \,
    \left( { f_{jk}(w_s)}^2 + { f_{kj}(w_s)}^2 + 2 { f_{jj}(w_s)} { f_{kk}(w_s)} \right),  \\
    \label{eq:Gp}
    &  G_{2p} & = M^{-1} \sum_{s=1}^M {(\theta - w_s)}^{2p} {\cal R } [f_{jk} (w_s)],
\end{eqnarray}
 and   $c_{2 \ell} = M^{-1} \sum_{s=1}^M {( \theta - w_s )}^{2 \ell}$.

\begin{proposition}
\label{prop:local-spec-biasandvar}
      Suppose $\{ X_t \}$ is a strictly stationary  Gaussian time series satisfying the Assumption A(4).  Then
\[
  \mbox{Var} [ \tilde f_{jk}  (\theta) ]  =   \delta^{-1}  n^{-1} \,
  \frac{ c_4^2 F_0 - 2 c_4 c_2 F_2 + c_2^2 F_4 }{ 4 {(c_4 - c_2^2)}^2 }
  + o(n^{-1})
  \]
   and 
\[
\mbox{Bias} [ \tilde f_{jk} (\theta) ] =  \frac{ c_4 G_0 - c_2 G_2 }{ c_4 - c_2^2 } -  f_{jk}(\theta)    + O(n^{-1})
  \]  
 as $n \tends \infty$ for a fixed  $\delta $. 
\end{proposition}
\vskip .1in
\noindent
{\bf  Proof:}    We focus on the case of $\theta = 0$, as $\theta = \pm \pi$ is similar with some
 notation changes.   Note that Assumption A(4) implies the autocovariances satisfy the summability condition
(B1) of Taniguchi and Kakizawa (2000), and the higher order autocumulants are zero because
 the process is Gaussian; hence Assumption B holds, and we can apply arguments from the proof of 
  Theorem \ref{thm:OLS-consistent} to conclude that 
 \begin{eqnarray*}
&  E [ \tilde{f}_{jk} (\theta) ]  & = O(n^{-1}) +  M^{-1} \sum_{s=1}^M \mbox{tr} ( \varphi (w_s) \, f (w_s) ) \\
&  &  = O(n^{-1}) + \frac{1}{2M} \sum_{s=1}^M  g(w_s) ( f_{jk} (w_s) + f_{kj} (w_s)) \\
&  &  = O(n^{-1}) + \frac{1}{M} \sum_{s=1}^M  g(w_s) {\cal R} [f_{jk} (w_s) ]. 
 \end{eqnarray*}
Using the definition of $G_{2p}$ given in (\ref{eq:Gp}), and $g(w) = (c_4 - c_2 w^2)/(c_4 - c_2^2)$,
 we obtain $E [ \tilde{f}_{jk} (\theta) ]   = O(n^{-1}) + (c_4 G_0 - c_2 G_2)/(c_4 - c_2^2)$, which
  yields the stated bias result.   
Adapting Lemma 3.1.1 of Taniguchi and Kakizawa (2000) to the case where
  frequencies $w_s \in [0, \pi]$, and using the fact that the process is Gaussian,
 the asymptotic variance of $ M^{-1} \sum_{s=1}^M \mbox{tr} ( \varphi (w_s) \, I (w_s) )$
 is given by
%
%
\[
  \frac{1}{ M^2} \sum_{s=1}^M \mbox{tr} ( f (w_s) \varphi (w_s) f (w_s) \varphi (w_s) ) 
  = \frac{1}{4 M^2} \sum_{s=1}^M  { g(w_s) }^2 \, \left(  f_{jk}^2 (w_s) + f_{kj}^2 (w_s) 
    + 2 f_{jj} (w_s) f_{kk} (w_s) \right),
 \]
 ignoring  terms that are  $o(n^{-1})$. 
 Expanding the square of $g(w_s)$ and using 
  (\ref{eq:Fp}), we obtain the stated variance expression since $M \approx \delta n$.  q.e.d.


  \vskip .1in
 
  The Mean Squared Error (MSE) of $\tilde f_{jk}  (\theta)$ equals the squared bias plus the variance.
The value of $\delta$ that minimizes the   MSE  can be calculated numerically, given 
the quantities $F_0$, $F_2$, $F_4$, $G_0$,  $G_2$, and $  f_{jk}  (\theta) $.  However, these
 quantities are unknown in practice; a pilot estimator, such as the flat-top
 spectral estimator of Politis (2011), can be used to estimate them -- yielding   
a data-based estimate of the optimal value of $\delta$ by plugging the  pilot estimators into the MSE formula.
  Such an empirically estimated $\delta$,
 which may be depend on $j$ and $k$,   will be denoted by $\widehat{\delta}_{jk}$;
  see Section \ref{sec.practical} for     explicit details.

\subsection{Fitting via Weighted Least Squares (WLS)}
 
   Consider a positive kernel function $K$ with domain $[0,1]$; 
usually the kernel is taken to be symmetric
about zero, but we focus on one-sided kernels due to our
 analysis at the boundaries of the frequency domain.   Define 
   $K_{\delta} (x) = {(2 \pi \delta)}^{-1} K (x/ (2 \pi \delta))$, so that
  $K_{\delta}$ becomes a function on $[0, \pi]$.  Then, we can define 
WLS weights   via $K_{\delta} ( \theta - w_s)$ for $1 \leq s \leq M$, 
  and we let $ {\bf W}$ be a diagonal matrix
with diagonal entries
given by $K_{\delta} ( \theta - w_s)$ for $  s =1,\ldots, M$.
  The WLS estimator of $ f_{jk}(\theta)$ can be obtained based on the matrix
  expression $ {( {\bf X}^{\prime} {\bf W} {\bf X} )}^{-1} {\bf X}^{\prime} {\bf W} {\bf Y}$,
   where ${\bf X}$ and ${\bf Y}$ have been defined in the previous section.
Note that the OLS is a special case of WLS in the case of equal weights, i.e.~letting  $K(x)=2 \pi \delta$.

 \begin{theorem} 
  \label{thm:WLS-consistent}
Assume Assumptions A(4) and B.  
Let  $ \tilde   f_{jk}(\theta)=  {[1,0]}^{\prime}
  {( {\bf X}^{\prime} {\bf W} {\bf X} )}^{-1} {\bf X}^{\prime} {\bf W} {\bf Y}$.  
Then,  $$  \tilde   f_{jk}(\theta) \convinp   f_{jk}(\theta) $$
as $M\to \infty$ as $n\to \infty$ but with
$M/n \to 0$.
\end{theorem} 
\vskip .1in
\noindent
{\bf  Proof:}  The proof follows the same structure as the proof of
Theorem  \ref{thm:OLS-consistent}; we provide a few details for the $\theta = 0$ case.
  Setting $\tilde{c}_{2 \ell} = M^{-1}  \sum_{s=1}^M K_{\delta} (w_s) w_s^{2 \ell}$,
  the weighting function for WLS is now
  $\tilde{g} (w) = (\tilde{c}_4 - \tilde{c}_2 w^2)/( \tilde{c}_4 \tilde{c}_0 - \tilde{c}_2^2)$.
  The rest of the proof is the same with $\tilde{g}$ in lieu of $g$, and we find that
  $\lim_{n \tends \infty} \tilde{g}( 2 \pi \delta x) =
   {(2 \pi \delta)} (K_{(4)} - K_{(2)} x^2)/ (K_{(4)} K_{(0)} - K_{(2)}^2 )$, since
   $\tilde{c}_{2 \ell} \tends {(2 \pi \delta)}^{2 \ell -1} K_{(2 \ell)}$ as $n \tends \infty$,
    where  $K_{(2 \ell)} = \int_0^1 K (  u) u^{2 \ell} du$.
   Then we update the expression for $f^{\delta}_{jk} (\theta)$ with
   \[
     2 \pi \delta  \,  \int_0^1  \left( \frac{ K_{(4)} - K_{(2)} x^2 }{ K_{(4)} K_{(0)} - K_{(2)}^2 }    \right) \,
    K_{\delta} (2 \pi \delta x) (f_{jk} (2 \pi \delta x) + f_{kj} (2 \pi \delta x)/2 \, dx,    
    \]
    which tends to $f_{jk} (\theta)$ as $\delta \tends 0$ using Assumption A(4).
    This completes the proof.  q.e.d.

  \vskip .1in

As before,
  asymptotic bias and variance expressions akin to the OLS case can be worked out,
which  now  depend upon the function $K$. However, we now need to 
allow for $\delta \tends 0$ in order to simplify the weights. To see why, 
  let $K_{(s)} = \int_0^1 K (  u) u^s du$ and
    $\widetilde{K}_{(s)} = \int_0^1 {[ K(u)]}^2 \, u^s \, du$. 
For the case $\theta = 0$    we obtain
\[
 M^{-1}  {\bf X}^{\prime} {\bf W} {\bf X}   \approx
     \left[ \begin{array}{cc}   {(2 \pi \delta)}^{-1} K_{(0)} &   {(2 \pi \delta)} K_{(2)} \\
 	  {(2 \pi \delta)}  K_{(2)} &   {(2 \pi \delta)}^{3} K_{(4)}  \end{array} \right];
\]
a similar approximation can be worked out for the case $\theta = \pi .$

 \begin{proposition}
\label{prop:wls-spec-biasandvar}
     Suppose $\{ X_t \}$ is a strictly stationary  Gaussian time series satisfying   Assumption A(4).      Then, 
\begin{eqnarray*}
 &  \mbox{Var} [ \tilde   f_{jk} (\theta) ] & =   O(\delta^{-2} n^{-2})   + o(n^{-1}) \\
  & & \; + \delta^{-1} n^{-1}   
    \left( { f_{jk}(\theta)}^2 + { f_{kj}(\theta)}^2 + 2 { f_{jj}(\theta)} { f_{kk}(\theta)} \right)    \,  
    \frac{  K^2_{(4)} \widetilde{K}_{(0)}  - 2   K_{(4)} K_{(2)} \widetilde{K}_{(2)} + 
     K_{(2)}^2 \widetilde{K}_{(4)}  }{ {(K_{(0)} K_{(4)} - K_{(2)}^2 )}^2 }  
 \end{eqnarray*}
  and
\[
  \mbox{Bias} [ \tilde   f_{jk}(\theta) ]  =   \frac{  { [ {\cal R }  f_{jk} ]}^{(4)} (\theta) }{ 24}  {( 2 \pi \delta)}^4 
    \,  \left( \frac{ K_{(4)}^2 - K_{(6)} K_{(2)} }{ K_{(4)} K_{(0)} - K_{(2)}^2 } \right)   +  o( \delta^4) + O(n^{-1})
  \]
 when $\delta \tends 0$  as $n\to \infty$ in such a way that  $\delta n\to \infty$.
\end{proposition}
\vskip .1in
\noindent
{\bf  Proof:}    Following the proofs of  Theorem \ref{thm:OLS-consistent} and Proposition
 \ref{prop:local-spec-biasandvar},  the estimator can be written as
  $\tilde{f}_{jk} (\theta) = M^{-1} \sum_{s=1}^M \mbox{tr} ( \varphi (w_s) I(w_s) )$
  up to $O_P (n^{-1})$ terms, where $\varphi (w) = .5 \tilde{g}(w) K_{\delta} (w)
    (e_k e_j^{\prime} + e_j e_k^{\prime})$
  and   $\tilde{g} (w) = (\tilde{c}_4 - \tilde{c}_2 w^2)/( \tilde{c}_4 \tilde{c}_0 - \tilde{c}_2^2)$.  
  Recall that  $\tilde{c}_{2 \ell} = M^{-1}  \sum_{s=1}^M K_{\delta} (w_s) w_s^{2 \ell}$,
  and as $n \tends \infty$ this has a limit of ${(2 \pi \delta)}^{2 \ell -1} K_{(2 \ell)}$.  
  Then $E [ \tilde{f}_{jk} (\theta) ]   = O(n^{-1}) +
   (\tilde{c}_4 \tilde{G}_0 - \tilde{c}_2 \tilde{G}_2)/(\tilde{c}_4 \tilde{c}_0- \tilde{c}_2^2)$,
  where $\tilde{G}_{2 p} = M^{-1} \sum_{s=1}^M {( \theta - w_s)}^{2p} K_{\delta} (w_s) 
   {\cal R } [f_{jk} (w_s)]$.   Since $ {\cal R } [f_{jk} (w)] =    \gamma_{jk} (0) + \sum_{h \geq 1} ( \gamma_{jk} (h) + \gamma_{jk}(-h) ) \cos (h w)$,  this is a real even function of $w$, which we denote by $r_{jk} (w)$ for short. 
    Then via Assumption A(4) we have a Taylor expansion around $\theta = 0$:
\[
 r_{jk} (w) = r_{jk} (0) + \frac{1}{2} r_{jk}^{(2)} (0) w^2 + \frac{1}{24} r_{jk}^{(4)} (0) w^4 + o(w^4),
\]
   where $r_{jk}^{(2)} (0)$ and $r_{jk}^{(4)} (0)$ denote the second and fourth derivatives of $r_{jk}$ at zero,
   respectively.   Plugging this into the formula for $\tilde{G}_{2p}$, and using
   \[
   M^{-1} \sum_{s=1}^M w_s^{2 \ell} K_{\delta} (w_s) = { (2 \pi \delta)}^{2 \ell -1} M^{-1} \sum_{s=1}^M
    {(s/M)}^{2 \ell} K (s/M) \tends { (2 \pi \delta)}^{2 \ell -1} K_{(2 \ell)},   
   \]
   we find that up to $o( \delta^{2p+3})$ terms    $\tilde{G}_{2p}$ is asymptotically given by
   \[
   { (2 \pi \delta)}^{2 p -1} K_{(2p)}  f_{jk} (0) + \frac{1}{2} { (2 \pi \delta)}^{2 p +1} K_{(2p+2)} 
    r_{jk}^{(2)} (0) + \frac{1}{24} { (2 \pi \delta)}^{2 p +3} K_{(2p+4)} r_{jk}^{(4)} (0).
  \]
  Therefore $E [ \tilde{f}_{jk} (\theta) ] $, up to $O(n^{-1})$ and $o( \delta^4)$ terms, is
\[
 f_{jk} (0) + \frac{1}{24} r_{jk}^{(4)} (0) {(2 \pi \delta)}^4 \left( \frac{ K_{(4)}^2 - K_{(2)} K_{(6)} }{
   K_{(4)} K_{(0)} -    K_{(2)}^2  } \right),
 \]  
  which yields the stated result for the bias.   The variance calculation follows the same argument
  used in the proof of Proposition \ref{prop:local-spec-biasandvar}, yielding (up to terms $o(n^{-1})$)
\[  
  \frac{1}{4 M^2} \sum_{s=1}^M  { \tilde{g}(w_s) }^2 \,  { K_{\delta} (w_s) }^2 \,  
  \left(  f_{jk}^2 (w_s) + f_{kj}^2 (w_s) 
    + 2 f_{jj} (w_s) f_{kk} (w_s) \right). 
  \]  
  By Assumption A(4) we can replace $  f_{jk}^2 (w_s) + f_{kj}^2 (w_s) 
    + 2 f_{jj} (w_s) f_{kk} (w_s)$ by its value at $\theta = 0$, and take it out of the summation
    at the cost of adding error terms that are $O(M^{-2})$.  The remaining summation can
     then be approximated using 
    \[
    M^{-1} \sum_{s=1}^M w_s^{2 \ell} K_{\delta}^2 (w_s) = { (2 \pi \delta)}^{2 \ell -2} M^{-1} \sum_{s=1}^M
    {(s/M)}^{2 \ell} {K (s/M) }^2  \tends { (2 \pi \delta)}^{2 \ell -2} \tilde{K}_{(2 \ell)},
  \]
  and hence the stated variance expression is obtained.  q.e.d.
  
  \vskip .1in


Proposition \ref{prop:wls-spec-biasandvar} quantifies the advantage of 
the local quadratic regression at the boundaries as the 
$\mbox{Bias} [ \tilde   f_{jk}(\theta) ]  $ becomes of order 
$\delta ^4$. Notably,   traditional methods of estimating 
$    f_{jk}(\theta) $  will have bias that is $O(\delta ^2)$
 under Assumption A(4); these include all 2nd order kernels employed
by Andrews (1991) but not the Bartlett kernel used by Newey and West
(1987), whose   bias  is $O(\delta  )$. 
Interestingly, the flat-top kernel estimator proposed by Politis (2011)
will also have bias that is $O(\delta ^4)$  under Assumption A(4), 
but the constant of proportionality is not tractable so as to afford us an
analytic comparison; an empirical comparison is carried out in Section \ref{sec.sims}.
We can summarize these results in the  following corollary.

\begin{corollary} 
\label{co:4.1}
 Suppose $\{ X_t \}$ is a strictly stationary  Gaussian time series satisfying   Assumption A(4).      Then, 
\[ \mbox{MSE} [ \tilde   f_{jk}(\theta) ]  = O(\delta ^8)+O(\delta^{-1}n^{-1}),
\]
whose order of magnitude is minimized when
 $\delta \sim C n ^{-1/9}$ for some constant $C>0$. 
Using such a choice of  $\delta$, the matrix estimator satisfies 
\begin{equation}
\label{eq.fromco:4.1}
\tilde   f(\theta)= f(\theta)+O_P(n ^{-4/9}) .
\end{equation} 
  \end{corollary}

\section{Practical implementation}
\label{sec.practical}

  \subsection{Data-based bandwidth choice}
  As in all nonparametric work, the choice of the bandwidth fraction 
   $\delta$ has a strong impact on finite-sample results, and   influences the asymptotic performance as well. 
  McElroy and Politis (2022) proposed a data-driven technique for empirically selecting $\delta$ in the univariate case
  by minimizing the MSE.  Since their method (which is the case $j=k$ of the present treatment)
  only depends upon the univariate spectral density being real and even (and not
necessarily non-negative), we can apply the
  same technique to the real portion of each cross-spectral density
$f_{jk}$.    This application yields a data-based estimator $\widehat{\delta}_{jk}$ for each
$\tilde  f_{jk} (\theta)$, by minimizing with respect to $\delta$ the MSE arising from
the bias and variance expressions in Proposition  \ref{prop:local-spec-biasandvar};
 we can collect them in a matrix $\widehat{\delta}$
whose  $  j,k  $ entry is $\widehat{\delta}_{jk}$.

Therefore, in order to compute the MSE
 we need to evaluate expressions (\ref{eq:Fp}) and (\ref{eq:Gp}) respectively.  Letting $d_{jk} (w)$ denote the
  determinant of the 2-by-2 spectral density matrix for the $j$th and $k$th series
  (i.e., $d_{jk} (w) = f_{jj} (w) f_{kk} (w) - f_{jk}(w) f_{kj} (w)$), and noting that
   this function --- as well as $    {\cal R } [f_{jk} (w)] = (f_{jk} (w) + f_{kj} (w))/2$ --- is real-valued,
   it is a simple matter to evaluate expression (\ref{eq:Fp}) without any complex arithmetic,
   as ${ f_{jk}(w)}^2 + { f_{kj}(w)}^2 + 2 { f_{jj}(w)} { f_{kk}(w)} = 
    2 d_{jk} (w) + 4  { ( {\cal R } [ f_{jk} (w)] )}^2$.  The real and imaginary parts of $f_{jk} (w)$ are obtained
    from the cross-covariances via
\begin{eqnarray*}
  & {\cal R } [f_{jk} (w)]  & = \gamma_{jk} (0) + \sum_{h \geq 1} ( \gamma_{jk} (h) + \gamma_{jk}(-h) ) \cos (h w) \\
 &  {\cal I } [f_{jk} (w)]  & =  \sum_{h \geq 1} ( \gamma_{jk} (h) - \gamma_{jk}(-h) ) \sin (h w).
  \end{eqnarray*}
  Similarly, the first formula
   can be applied to compute $f_{jj} (w)$ and $f_{kk} (w)$, since these are real;
  thence we obtain 
  \[
  d_{jk} (w) = f_{jj} (w) f_{kk} (w) - { [{\cal R } f_{jk} (w)] }^2 - { [{\cal I } f_{jk} (w)] }^2,
  \]
  since $f_{jk} (w) f_{kj } (w) = {| f_{jk} (w) |}^2 = { [{\cal R } f_{jk} (w)] }^2 + { [{\cal I } f_{jk} (w)] }^2$.

  As argued in Politis (2011),  it is natural to allow bandwidth fraction to be customized/optimized to each entry of the spectral density 
  matrix,  since the degree of autocorrelation and cross-correlation present (which dictates the optimal selection of bandwidth) can 
  vary among components. 
In other words, our matrix estimator $\tilde  f  (\theta)$ has 
$  j,k  $ entry $\tilde  f_{jk} (\theta)$  that is computed using its own
optimal data-based bandwidth fraction $\widehat{\delta}_{jk}$.
 In practice this is extremely important, since we have found that  $\widehat{\delta}_{jk}$ can vary widely
 for various $j,k$.  
 
 \begin{Remark} \rm
  The numerical optimization required to find the minimizing $\widehat{\delta}_{jk}$  requires some care,
  because MSE as a function of $\delta$ can have an oscillatory pattern -- inherited from sinusoidal
  terms of high frequency in the spectrum or cospectrum.   Finding a local minimum 
  can lead to an inappropriate choice of $\widehat{\delta}_{jk}$, and substantially more bias than 
  would arise with the global minimum.  We use a golden section search algorithm in our own
  implementation, but recommend a  complete grid search if resources permit. 
 \end{Remark}

\subsection{Correction towards positive definiteness}
Corollary \ref{co:4.1} shows that  the matrix estimator
$\tilde  f  (\theta)$ converges to its 
target $  f  (\theta)$  at a fast  rate. Recall that $f(w)$ is 
  non-negative definite for all $w$. 
 If  $  f  (\theta)$  is positive
definite (as it will typically be when $ \theta =0$),
 it follows that $\tilde  f  (\theta)$
will eventually (for large $n$) be positive
definite as well. However, there is no guarantee that  $\tilde  f  (\theta)$
will be non-negative definite (let alone positive
definite) in finite samples. 
A correction towards positive definiteness is therefore in order, as
proposed by Politis (2011), and further developed by McMurry and Politis (2015).

There are several ways to correct $\tilde  f  (\theta)$
  towards positive definiteness. The simplest way is to use its 
eigenvalue decomposition 
$\tilde  f  (\theta)= U D U^\prime$, 
where $U$ is a unitary matrix, and $D$ is diagonal with 
$jj$ element $d_j$. Recall that $\tilde  f  (\theta)$ is Hermitian, and therefore
the eigenvalues $d_j$ are real; the question is whether they are all 
positive (or at least non-negative). 

For this reason, we can start by modifying the eigenvalues.
Let $d_j^+ = \max ( d_j, 0)$, and $d_j^{(\epsilon)} = \max ( d_j, 
\epsilon/n)$ for some small $\epsilon >0$. 
We now define the matrices
\begin{equation}
  \tilde  f ^+ (\theta)= U D^+ U^\prime \ \ \mbox{and} \ \ 
\tilde  f ^{(\epsilon)} (\theta)= U D^{(\epsilon)} U^\prime, 
\label{eq.plus}
\end{equation}
where $D^+$  and $D^{(\epsilon)}$ are diagonal with 
$jj$ element $d_j^+$ and $ d_j^{(\epsilon)} $ respectively.
The matrix $\tilde  f ^+ (\theta)$  is non-negative definite, and $
\tilde  f ^{(\epsilon)} (\theta)$ is positive definite by construction. 
However, as shown by Politis (2011), these two matrices share the
same fast rate of convergence with 
 $\tilde  f  (\theta)$, i.e., equation (\ref{eq.fromco:4.1}) from
 Corollary \ref{co:4.1}  holds true with
either $\tilde  f ^+ (\theta)$  or $
\tilde  f ^{(\epsilon)} (\theta)$  in place of $\tilde  f   (\theta)$.
 
\begin{Remark}
\label{Re:5.2}  \rm
The above idea works well in practice if the  multivariate  time series $\{ X_t \}$ 
is comprised of univariate time series that have approximately the same scale, i.e.,   
 the diagonal elements of $\gamma (0)$ are of the same order of magnitude. 
If not, the following modification may be helpful; we state it in connection to our original problem, viz. estimation of the long-run covariance 
matrix $\Omega  $ that appears in the CLT 
based on data $X_1,\ldots, X_n$ as given in eq. (\ref{eq.CLT}). 
\end{Remark} 

\begin{enumerate}
\item
Define new data 
$Y_t= \hat A (X_t-\bar X_n) $ for $t=1,\ldots, n$,  where $ \hat A
$ is diagonal with  $jj$ element   equal to $1/\sqrt{\hat \gamma_{jj}(0)}$. 
\item Use local quadratic regression on the new data $Y_1,\ldots, Y_n$ and the 
correction of  the right-hand-side of eq. (\ref{eq.plus}) to estimate the
 spectral density matrix of $\{ Y_t \}$ at the origin; denote that estimator by $
\tilde  f _Y^{(\epsilon)} (0)$.
\item Note that eq. (\ref{eq.CLT}) and consistency of $\gamma_{jj}(0)$
imply that
\[
\sqrt{n}  \  \bar Y_n     \convinlaw N_m(0,A\Omega A ) \ \ \mbox{as} \ \ n\to \infty,
\]
where  $   A$ is a diagonal matrix
with  $jj$ element   equal to $1/\sqrt{  \gamma_{jj}(0)}$.
\item Since $2\pi \tilde  f _Y^{(\epsilon)} (0)$ is an estimate of  $A\Omega A$, 
we may finally estimate $\Omega$ by 
\[
\hat \Omega = 2\pi  \hat A ^{-1} \tilde  f _Y^{(\epsilon)} (0) \hat A ^{-1}.
\]
\end{enumerate}

\section{Simulations}
\label{sec.sims}
 
Recall that Politis (2011)
includes simulations showing the favorable performance
of the flat-top matrix estimators as compared to traditional estimators
using non-negative kernels. 
In order to compare to the numerical results
of  Politis (2011), we simulate the same  
data generating processes (DGP),  and focus on 
the boundary frequencies  $w = 0, \pm \pi$.

 The first DGP is defined as 
 \begin{eqnarray*}
   X_{t,1} & = & { (1 - .75 L) }^{-1} Z_{t,1} \\
    X_{t,2} & = & 2 (1 + L) Z_{t,2},
\end{eqnarray*}
 where $L$ is the lag operator and $\{ Z_t \}$ is a bivariate Gaussian white noise sequence with identity covariance matrix.
  Note that no cross-sectional dependence exists in this process, and hence the cross-spectral density is zero.  The spectral
  density matrix  is given by
  \begin{equation}
  \label{eq:dgp1.spec}
   f(w) = \frac{1}{2 \pi}  \, \left[ \begin{array}{cc}  {| 1 - .75 e^{-i w} | }^{-2} & 0 \\ 0 & 4 {| 1 + e^{-i w} | }^2 
   \end{array} \right].
\end{equation}
Evaluating at $w=0$, we see that $f(0) $ is an identity matrix scaled by $8/\pi$; in particular, the two univariate spectral densities
 have equal values at $w =0$, although their shape in a neighborhood of the origin is somewhat different.
 At frequency $w = \pi$, all entries of the spectral density matrix are zero except the first diagonal, which equals $8/49 \pi$.

   The second DGP is defined as 
 \begin{eqnarray*}
   X_{t,1} & = & (1 - L)  Z_{t,1} \\
   X_{t,2} & = & X_{t+7,1} + {(1 + .75 L)}^{-1} Z_{t,2},
\end{eqnarray*}
 where  $\{ Z_t \}$ is as above.
  Here, there exists cross-sectional dependence through the second component process
   depending upon a future value (seven steps ahead)
   of the first component process.   The spectral
  density matrix  is given by
  \begin{equation}
  \label{eq:dgp2.spec}
   f(w) = \frac{1}{2 \pi}  \, \left[ \begin{array}{cc}  {| 1 - e^{-i w} | }^{2} &  e^{-i 7 w}   {| 1 - e^{-i w} | }^{2}    \\ 
     e^{i 7 w}   {| 1 - e^{-i w} | }^{2}  &    {| 1 - e^{-i w} | }^{2} +  {| 1 + .75 e^{-i w} | }^{-2} 
   \end{array} \right].
\end{equation}
 The lag seven dependence is manifested in the off-diagonal entries, through the terms $e^{\pm i 7 w}$.
Evaluating at $w=0$ and $w = \pi$, we obtain
\[
  f(0) = \frac{1}{ 2 \pi} \, \left[ \begin{array}{cc} 0 & 0 \\ 0 & 16/49 \end{array} \right]
  \quad \mbox{and} \quad
   f(\pi) = \frac{1}{ 2 \pi} \, \left[ \begin{array}{cc} 4 & -4 \\ -4 & 20 \end{array} \right]
   \]
   respectively.   So although there is cross-sectional dependence between the series, this effect vanishes at frequency $w = 0$;
   the dependence is non-vanishing (and negative) at $w = \pi$.   
 
We simulated $10^4$ Monte Carlo replications of each DGP (with $n = 100, 500$), and compared the flat-top 
 estimator to the local quadratic regression technique (with $\widehat{\delta}$ determined empirically) as discussed
 in Sections \ref{sec.method} and \ref{sec.practical}.   The  flat-top 
 estimator employed the 
 infinitely differentiable flat-top taper (McMurry and Politis, 2004),  with  the $c_{ef}$ tuning parameter
 (Politis, 2011)    computed using  $\varepsilon = .01$.  Also, in the  algorithm used to compute the flat-top
  bandwidth, i.e., the Empirical Rule of Politis (2011),
we set $C_0 = 2$ and $K_n = 5$.
Both estimates  of $f (0)$ were 
 modified  (when necessary)  to be non-negative definite,
i.e., we employed the first of the two equations given in (\ref{eq.plus}).

\begin{table}[ht]
\centering
\begin{footnotesize}
\begin{tabular}{|cc|ccc|ccc|}
  \hline
 \multicolumn{2}{|c|}{First DGP} & \multicolumn{3}{c|}{local quadratic}  
   & \multicolumn{3}{c|}{flat-top}       \\  \hline \hline
 $n$  &  Component & Bias & SD & RMSE & Bias & SD & RMSE   \\ 
  \hline
  \multirow{3}{*}{ $100$ } & $f_{11} (0)$  & -0.839 &  1.129 &  1.407  &   -0.048  &  1.769 &  1.770  \\
  					& $f_{12} (0)$ &   -0.002  & 0.456 & 0.456 &  0.005  &  0.676  &  0.676  \\
  					& $f_{22} (0)$ &    -0.197  & 0.677 & 0.705  &   0.010 &  0.908 & 0.908  \\
\hline  
 \multirow{3}{*}{ $500$ } & $f_{11} (0)$  &   -0.275 &   0.758 & 0.807  &  0.000 &   0.926 &  0.926 \\
  					& $f_{12} (0)$ &    0.001 &  0.198 & 0.198 & 0.000  &  0.265 & 0.265 \\
  					& $f_{22} (0)$ &  -0.099 &  0.327 & 0.342 &  0.001 &  0.359 & 0.359 \\ 		 
\hline    
\end{tabular}
\end{footnotesize}
\caption{\baselineskip=10pt Bias, Standard Deviation, and RMSE  for spectral density estimators at frequency $\theta = 0$,
 for a bivariate Gaussian {} process with spectral density (\ref{eq:dgp1.spec}). 
  Sample size is $n= 100,  500$.  
  Flat-top tapered   estimation and local quadratic spectral estimation  is      considered 
  with estimated optimal window $\widehat{\delta}$.}
\label{table-main:dgp1w0}
\end{table}

\begin{table}[ht]
\centering
\begin{footnotesize}
\begin{tabular}{|cc|ccc|ccc|}
  \hline
 \multicolumn{2}{|c|}{First DGP} & \multicolumn{3}{c|}{local quadratic}  
   & \multicolumn{3}{c|}{flat-top}       \\  \hline \hline
 $n$  &  Component & Bias & SD & RMSE & Bias & SD & RMSE   \\ 
  \hline
  \multirow{3}{*}{ $100$ } & $f_{11} (\pi)$  &   -0.001 &  0.025 & 0.025 &  0.008  & 0.036 & 0.037  \\
  					& $f_{12} (\pi)$ &   0.000  &  0.018 & 0.018  &  0.000 &  0.021 & 0.021  \\
  					& $f_{22} (\pi)$ &  0.016 &  0.021 & 0.026  &   0.042 &  0.053 & 0.068  \\
\hline  
 \multirow{3}{*}{ $500$ } & $f_{11} (\pi)$  &  -0.001 &   0.014 & 0.014  &  0.001  &  0.017 & 0.017  \\
  					& $f_{12} (\pi)$ &   0.000  &  0.008  & 0.008 &  0.000  &  0.011 & 0.011  \\
  					& $f_{22} (\pi)$ &   0.004  &  0.004 & 0.006 &  0.018  &  0.024 & 0.030 \\ 		 
\hline    
\end{tabular}
\end{footnotesize}
\caption{\baselineskip=10pt Bias, Standard Deviation, and RMSE  for spectral density estimators at frequency $\theta = \pi$,
 for a bivariate Gaussian {} process with spectral density (\ref{eq:dgp1.spec}). 
  Sample size is $n= 100,  500$.  
  Flat-top tapered   estimation and local quadratic spectral estimation  is      considered 
  with estimated optimal window $\widehat{\delta}$.}
\label{table-main:dgp1wpi}
\end{table}

 For the first DGP---where there is no cross-correlation---we see that the local quadratic estimator  
 improves upon the flat-top almost uniformly, over all
matrix entries and either of the boundary  frequencies  $w = 0,  \pi$;
see Tables 1 and 2. 
Although the local quadratic  has a higher bias than the flat-top, this
 is more than
 compensated by the lower variability, leading to a lower
RMSE (square root of MSE).  As expected, both methods have improved performance as
 $n$ increases from $100$ to $500$;  it appears that the  
improvement offered by the   local quadratic is more pronounced 
   at the smaller sample size.

  For the second DGP---where there exists a non-trivial cross-spectrum---when $\theta = 0$,   
 the local quadratic has   superior RMSE performance 
for  $f_{11} (0)$ and $f_{22} (0)$; see Table 3. 
The two methods perform comparably for $f_{12} (0)$
when $n=100$, while the flat-top  gets a small edge when the
sample is increased. 
Moreover, in the   case $\theta = \pi$  Table 4 shows 
a  similar superiority of the local quadratic 
 over the flat-top---except for the case of $f_{21} (\pi)$ with  $n = 500$, where results are identical.

\begin{table}[ht]
\centering
\begin{footnotesize}
\begin{tabular}{|cc|ccc|ccc|}
  \hline
 \multicolumn{2}{|c|}{Second DGP} & \multicolumn{3}{c|}{local quadratic}  
   & \multicolumn{3}{c|}{flat-top}       \\  \hline \hline
 $n$  &  Component & Bias & SD & RMSE & Bias & SD & RMSE   \\ 
  \hline
  \multirow{3}{*}{ $100$ } & $f_{11} (0)$  &   0.007  & 0.008 & 0.010 &  0.012 &   0.014 &  0.018 \\  
  					& $f_{12} (0)$  &  -0.001 &  0.013 & 0.013 &  0.001  & 0.012 &  0.013 \\
  					& $f_{22} (0)$ &  0.008 &  0.035 & 0.036 &  0.015 &  0.039 & 0.042 \\
  \hline  
 \multirow{3}{*}{ $500$ } & $f_{11} (0)$  &  0.001 &  0.001 & 0.002 &  0.005  & 0.006 &  0.008 \\
  					& $f_{12} (0)$ & 0.000  & 0.003 & 0.003 &  0.001 &  0.002 & 0.002 \\
  					& $f_{22} (0)$ &  0.001 &  0.015 & 0.015 &   0.001 &  0.015 & 0.015  \\
\hline    
\end{tabular}
\end{footnotesize}
\caption{\baselineskip=10pt Bias, Standard Deviation, and RMSE  for spectral density estimators at frequency $\theta = 0$,
 for a bivariate Gaussian {} process with spectral density (\ref{eq:dgp2.spec}). 
  Sample size is $n= 100,  500$.  
  Flat-top tapered   estimation and local quadratic spectral estimation  is      considered 
  with estimated optimal window $\widehat{\delta}$.}
\label{table-main:dgp2w0}
\end{table}

\begin{table}[ht]
\centering
\begin{footnotesize}
\begin{tabular}{|cc|ccc|ccc|}
  \hline
 \multicolumn{2}{|c|}{Second DGP} & \multicolumn{3}{c|}{local quadratic}  
   & \multicolumn{3}{c|}{flat-top}       \\  \hline \hline
 $n$  &  Component & Bias & SD & RMSE & Bias & SD & RMSE   \\ 
  \hline
  \multirow{3}{*}{ $100$ } & $f_{11} (\pi)$  &    -0.027 &  0.221 & 0.223  &   0.030 &   0.262 & 0.264  \\  
  					& $f_{12} (\pi)$  &   0.283  & 0.610 & 0.673  &  0.133 &  0.734 & 0.746  \\
  					& $f_{22} (\pi)$ &   -0.910 &   1.567 & 1.812 &  -0.337 &  2.104 & 2.131 \\
  \hline  
 \multirow{3}{*}{ $500$ } & $f_{11} (\pi)$  &  -0.024 &  0.083 & 0.086  &   0.001 &  0.092 & 0.092  \\
  					& $f_{12} (\pi)$ &  0.110 &  0.336 & 0.353 &  0.003 &  0.353 & 0.353 \\
  					& $f_{22} (\pi)$ &  -0.433 &  0.880 & 0.981  &  -0.158 &  1.015 & 1.027   \\
\hline    
\end{tabular}
\end{footnotesize}
\caption{\baselineskip=10pt Bias, Standard Deviation, and RMSE  for spectral density estimators at frequency $\theta = \pi$,
 for a bivariate Gaussian {} process with spectral density (\ref{eq:dgp2.spec}). 
  Sample size is $n= 100,  500$.  
  Flat-top tapered   estimation and local quadratic spectral estimation  is      considered 
  with estimated optimal window $\widehat{\delta}$.}
\label{table-main:dgp2wpi}
\end{table}

\section{Real Data Application}
\label{sec.data}

 Soaring inflation is of widespread interest, and the linkages to the unemployment rate are subtle;
 some may assert these are necessarily inversely correlated, but there exist counter-examples,
 such as the post Great Recession epoch of low inflation and low unemployment.  
 We apply this paper's method to recent estimates of mean inflation and unemployment
 in order to highlight the possibilities for the proposed methodology,  without seeking to wade
 into a contentious and nuanced econometric topic that has numerous public policy implications.

 To this  end---of providing an illustration---we study the U.S.  Consumer Price Index
  (CPI)\footnote{U.S. Bureau of Labor Statistics, 
   Consumer Price Index for All Urban Consumers:
   All Items in U.S. City Average [CPIAUCSL], retrieved from FRED, 
   Federal Reserve Bank of St. Louis on
October 24, 2022.} and
   the U.S.   Unemployment Rate (UR)\footnote{U.S. Bureau of Labor Statistics, 
   Unemployment Rate [UNRATE], retrieved from FRED, 
   Federal Reserve Bank of St. Louis on
October 24, 2022.}.
  The common span for the two monthly time series is January 1948 through September 2022;
  applying a log difference to CPI to obtain its growth rate
  (and multiplying by  $12*100$ to convert CPI growth to  an annual percentage), we compose the two variables
  into a bivariate  time series  $\{ X_t \}$.   Note that both components of $X_t$ are on the same scale,
   so the possible  re-scaling suggested in Remark \ref{Re:5.2} is not necessary. 

   We wish to examine a more recent period,
  wherein a stationary hypothesis is tenable; therefore we exclude the Covid-19 epoch,
  and consider CPI growth and UR for January 2001 through December 2019, yielding $n = 240$.
  Over this period the sample means for CPI growth and UR are $ 2.126 \%$ and $ 5.881 \%$, respectively.  
  Here, we investigate whether the inflation rate and unemployment rate are   significantly   different from $3 \%$ and $5 \%$, respectively.
    These thresholds are chosen by the following reasoning: either a UR higher than $5 \%$
    or an inflation rate higher than $3 \%$ may be cause for concern,  whereas lower values
    might be viewed positively---although some degree of
    unemployment  (among those persons in the labor force) and inflation 
     is typically present (due to churn) even  in a healthy labor market,
    as desirable workers move freely between opportunities.
    
    A null hypothesis for the bivariate mean $\mu$ can be tested by the bivariate sample mean $\bar{X}_n$, which
     under broad conditions is asymptotically normal with asymptotic covariance matrix $2 \pi f(0)/n$; see eq.~(\ref{eq.CLT}). Hence, to test  $  H_0 : \mu = \mu_0$  one can use the Wald statistic
\[
 \frac{  n } {2 \pi} \,  {( \bar{X}_n - \mu_0 )}^{\prime}  \, {f (0)}^{-1} \, (\bar{X}_n - \mu_0 ), 
\]
 where $\mu_0 = {[3 \%,5 \%]}^{\prime}$.
As usual,  the Wald test involves  the 2-sided alternative
  $\mu \neq \mu_0$, so it will reject $  H_0$ 
if either   UR and/or inflation differ from their respective threshold. 

 Under $H_0$, the Wald statistic has an asymptotic $\chi^2_2$ distribution.  Since $f(0)$ is unknown, it is natural to estimate it  
  and compute the corresponding Wald statistic
\begin{equation}
\label{eq:Wald}
 \frac{  n } {2 \pi} \,  {( \bar{X}_n - \mu_0 )}^{\prime}  \, { \widehat{f}(0)}^{-1} \, (\bar{X}_n - \mu_0 )  .
\end{equation}
By Slutsky's Lemma, if  $\widehat{f}(0) $ is consistent for
$f(0)$,  the above practical Wald statistic will also 
have an asymptotic $\chi^2_2$ distribution under $H_0$.

 We adopt this approach, 
  using both the  infinitely differentiable  flat-top estimator 
and the local quadratic estimator.  Recall that in the previous 
section we had employed a non-negative definite modification 
of the resulting estimators. Here, it is important to strengthen this
to a positive definite
  modification  since we will be computing the matrix inverse, and any ill-conditioning in $\widehat{f} (0)$
  will cause gross numerical instabilities; 
  we used
  $\epsilon = .01$ in the construction given in eq.~(\ref{eq.plus}).
  However, in our results it happened that all estimates are positive definite.
The resulting  flat-top  and  local quadratic estimates
are  given by  
  \[
   \widehat{f} (0) = \left[ \begin{array}{cc}     
     42.17  &   112.23  \\   112.23  &  348.28  \end{array} \right]
   \qquad \mbox{and} \qquad
   \widetilde{f} (0) = \left[ \begin{array}{cc}     
     31.19  &   66.39  \\   66.39  &  248.55  \end{array} \right]    
\]
respectively.   Using the flat-top estimator, the Wald test statistic
(\ref{eq:Wald})  has a value of 
 $8.61$ with a p-value of $0.014$, indicating a strong rejection of the null.
 However, using the   local quadratic estimator we obtain
 a Wald statistic value of $3.61$ and a p-value of $0.164$, 
which does not lead to a rejection of $H_0$.

We remark that the
  estimate of $f_{21} (0)$ has a substantial impact
   on the Wald statistic. Even though the diagonal entries of the 
   flat-top estimate of $f (0)$ are larger than those of the 
   local quadratic, the Wald statistic for the former is larger due to
   the higher estimated coherence  $f_{21} (0)/\sqrt{f_{11} (0) f_{22} (0)}$. 
Hence, there is a genuinely multivariate phenomenon at work 
in this real data example.  

In view of these  vastly conflicting results the practitioner 
would be at a loss as to how to proceed. 
Nevertheless, our simulations of Section \ref{sec.sims}
coupled with the extensive (univariate) simulations of 
McElroy and Politis (2022) indicate that 
the local quadratic method of estimating
$f(0)$ is more accurate than the flat-top. 
Hence, the practitioner is advised to adopt the
local quadratic estimate of $f(0)$, in which case 
 the null  hypothesis is {\it not} rejected;  we conclude that the economy 
 is not out of balance  (with respect to inflation and employment).

\section*{Acknowledgments}

This report is released to inform interested parties of research and to encourage discussion.  The views expressed on
statistical issues are those of the authors and not  those of the U.S. Census Bureau.  Research of the second
 author partially   supported by NSF grant DMS 19-14556.
 


\begin{thebibliography}{999}

 \bibitem{andrews91}  Andrews, D. (1991). Heteroskedasticity and autocorrelation
consistent covariance matrix estimation, {\it Econometrica}, {\bf 59}, 817-858.




   \bibitem{andrews92}  Andrews, D. and Monahan, J. (1992). An improved
heteroskedasticity and autocorrelation
consistent covariance matrix estimator, {\it Econometrica}, {\bf 60}, 953-966.

\bibitem{B46}  Bartlett, M.S. (1946). On the theoretical specification of
 sampling  properties of autocorrelated time series, {\it J. Roy. Statist. Soc., Suppl.}, 8, 27-41.
 
\bibitem{} Billingsley, P. (1995).   {\it Probability and measure},  John Wiley \& Sons. 

 \bibitem{brill} Brillinger, D.R. (1981). 
  {\it Time Series: Data Analysis and Theory},
Holden-Day, New York.

 

   \bibitem{brock} Brockwell, P. J.  and Davis, R. A.
  (1991).  {\it Time Series: Theory and Methods, 2nd ed.},
Springer, New York.



\bibitem{D46} Daniell, P.J.  (1946).  Discussion of paper by M.S. Bartlett.
{\it J. Roy. Statist. Soc. Suppl.}, vol. 8, pp. 88-90.
  
 \bibitem{hamilton94}
Hamilton, J.D. (1994). {\it Time Series Analysis}. Princeton
University Press, Princeton.



   \bibitem{hannan} Hannan, E.J. (1970).
   {\it Multiple Time Series}, John Wiley, New York.

\bibitem{jones} Jones, C.I.  (2016).  
  The facts of economic growth. In  {\it Handbook of Macroeconomics} (Vol. 2, pp. 3-69). Elsevier.

\bibitem{} Kiefer,  N.  and  Vogelsang,  T.  (2002).
  Heteroscedastic-autocorrelation  robust  standard  errors
using the Bartlett kernel without truncation. {\it Econometrica}, 70, 2093-2095.

\bibitem{} Kiefer,   N.  and  Vogelsang,   T.  (2005).  A  new  asymptotic  theory  for
heteroscedasticity-autocorrelation robust tests. {\it Econometric Theory}, 21, 1130-1164.

\bibitem{}   McElroy, T.S. and   Politis, D.N. (2014). 
 Spectral density and spectral distribution inference for long memory time series via fixed-b asymptotics,
  {\it J. of Econometrics}, vol. 182, pp. 211-225.

\bibitem{}   McElroy, T.S. and   Politis, D.N. (2020). 
{\it Time Series: A First Course with Bootstrap Starter}, Chapman and Hall/CRC Press,  Boca Raton.
 
\bibitem{}   McElroy, T.S. and   Politis, D.N. (2022).  Estimating the spectral density at frequencies near zero.
Published online,  {\it Journal of the American Statistical Association}. 
arXiv preprint 2208.02300.

\bibitem{} McMurry, T. and Politis, D.N. (2004).     Nonparametric regression with infinite order flat-top kernels. 
{\it Journal of Nonparametric Statistics}, 16(3-4), pp.549-562.

\bibitem{} McMurry, T. and Politis, D.N. (2015).  High-dimensional autocovariance matrices and optimal linear prediction (with Discussion), {\it Electronic Journal of Statistics}, vol. 9, pp. 753-788.
 
   \bibitem{newey-west87} Newey, W. and West, K. (1987). A simple,
positive semi-definite, heteroskedasticity and autocorrelation
consistent covariance matrix,  {\it Econometrica}, {\bf 55}, 703-708.


   \bibitem{newey-west94} Newey, W. and West, K. (1994).
 Automatic lag selection in  covariance matrix estimation, {\it Rev. Econ. Studies},
 {\bf 61}, 631-653.

\bibitem{}
 Paparoditis, E.  and   Politis, D.N. (2012). Nonlinear spectral density 
 estimation: thresholding the correlogram, {\it J. Time Series Analysis}, vol. 33, no. 3, pp. 386-397.
 

\bibitem{} Parzen, E. (1957).
On consistent estimates of the spectrum of a stationary time series.
{\it Ann. Math. Statist.}, Vol. 28, No. 2., pp. 329-348. 


\bibitem{pz} Parzen, E. (1961).  Mathematical considerations in the estimation
of spectra, {\it Technometrics}, vol. 3,  167-190.

  \bibitem{dp2001} Politis, D.N. (2001). On nonparametric function estimation
 with infinite-order flat-top kernels, in {\it
Probability and Statistical Models with applications,} Ch. Charalambides et al. (Eds.), Chapman and Hall/CRC
Press,       Boca Raton,   pp. 469-483.

\bibitem{dp2003} Politis, D.N. (2003). Adaptive bandwidth choice,
{\it J. Nonparam. Statist.,} vol. 15, no. 4-5, 517-533.
 

\bibitem{dp2005} Politis, D.N. (2011). 
Higher-order accurate, positive semi-definite  estimation
 of large-sample  covariance and spectral density matrices,
 {\it Econometric Theory}, vol. 27, no. 4, pp. 703-744.


\bibitem{}  Politis, D.N. and  Romano, J.P. (1995).  Bias-corrected 
nonparametric spectral estimation, {\it  J. Time Ser. Analysis,} vol. 16, No. 1,  pp. 67-104. 


  \bibitem{ros} Rosenblatt, M. (1985).
{\it Stationary Sequences and Random Fields}, Birkh\"auser, Boston.

 

 
 
\bibitem{west} West, K.D. (1997).   Another heteroskedasticity--- 
and autocorrelation---consistent covariance matrix estimator,
{\it J. Econometrics}, 76, pp. 171-191.

 \bibitem{}   Wu, W.B. (2005). Nonlinear system theory: another look
 at dependence, {\it Proc. Nat. Acad. Sci.}, vol. 102, no. 40, pp. 14150--14154.


\end{thebibliography}
\end{document}